\documentclass[12pt,a4paper,colorlinks]{article}
\pdfoutput=1
\usepackage[cp1251]{inputenc}
\usepackage{fontenc}
\usepackage{cite}
\usepackage{amsmath}
\usepackage{amsfonts}
\usepackage{amssymb}
\usepackage{xcolor}
\usepackage{graphicx}

\usepackage{geometry} 
\graphicspath{{./img/}}
\begin{document}
\title{Analyticity and sum rules for photon GPDs}
\author{ I.R. Gabdrakhmanov, O.V. Teryaev}
\maketitle
\abstract{The photon is explored as an object to test the
applications of QCD to the perturbatively calculable collinear parton
distributions. We investigate analytic properties of DVCS amplitudes and related
sum rules of generalized parton distributions of the photon using as an input
their earlier calculations in the leading order. The relation of these GPDs to
the quintessential functions  in the framework of the dual parametrization approach is also found.}

\section{Introduction}
\label{intro}
\begin{figure}[h!]\centerline{
\begin{minipage}{0.4\linewidth}
\includegraphics[width=1\linewidth]{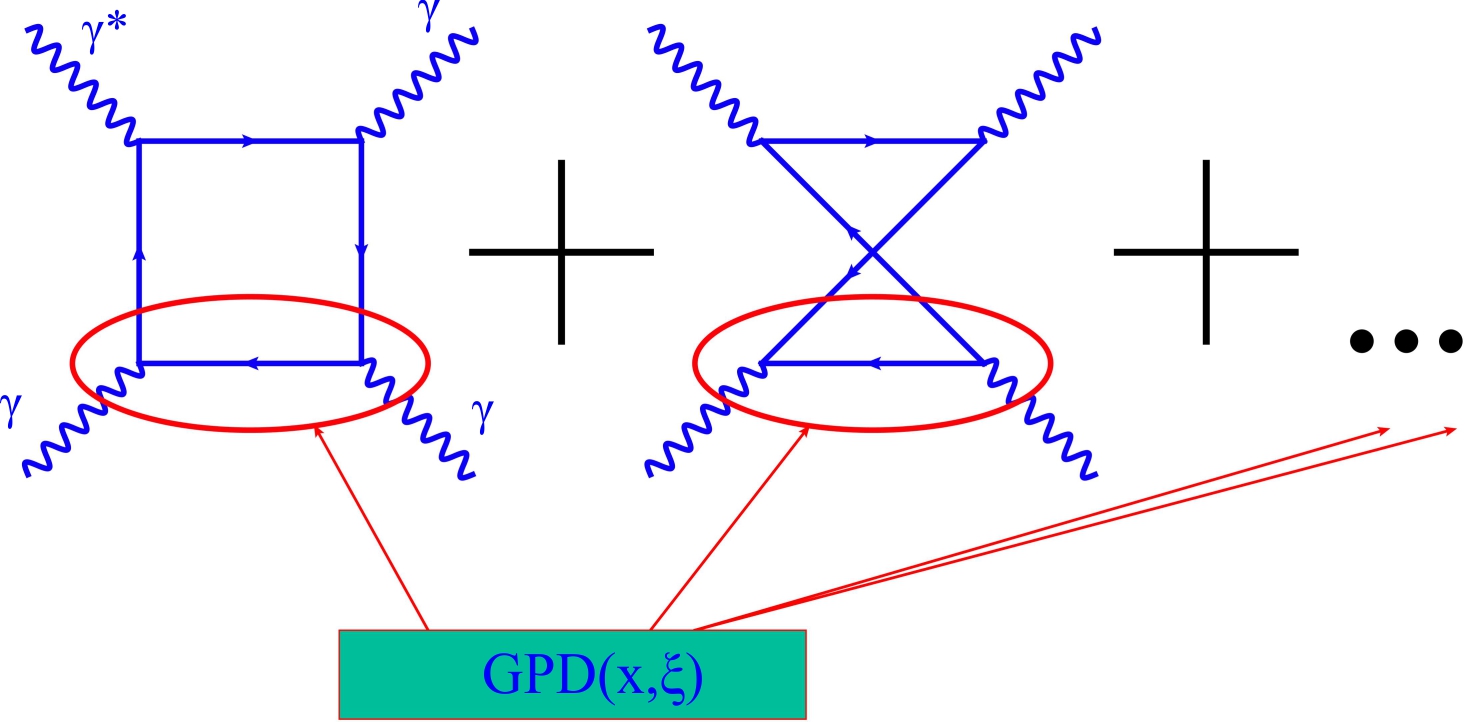}
\caption[]{$\gamma^* \gamma \rightarrow \gamma \gamma$ scattering}\label{ph_gpd}
\end{minipage}}
\end{figure}

In the concept of QCD factorization the amplitude of a hard process is expressed in terms of convolution of a hard perturbative amplitude and nonperturbative function characterizing nonperturbative strong interactions. In particular Generalized Parton Distributions (GPDs) and Generalized Distribution Amplitudes (GDAs) \cite{Belitsky:2005qn,Diehl:2003ny,Goeke:2001tz}
specifying low momentum scale strong interactions are used in the exclusive processes. In this Letter, we explore the application of the QCD tools to the photon. The unique feature of the photon is that in the leading order its partonic content can be calculated in perturbative QED \cite{Friot:2006mm,ElBeiyad:2008ss}.
Investigation of the partonic content of the photon (including the gluonic corrections) started with the seminal paper by Witten \cite{Witten:1977ju}. More recently the deeply virtual Compton scattering (DVCS) amplitude with the photon target (Fig.\ref{ph_gpd}) was calculated, and then GPD and GDA functions characterizing quark content of the photon were extracted. DVCS on a photon is therefore a kind of toy model for QCD which has been recently generalized for the case of impact parameter dependent GPDs \cite{Mukherjee:2011bn,Mukherjee:2011an}. This provides an incredible opportunity to test our current mathematical tools of GPDs studies.
Exploring the connection \cite{Teryaev:2001qm}
of usual GPDs, where skewness $|\xi|<1$, with the Generalized
Distribution Amplitudes (GDAa), where $|\xi|>1$, we checked
holographic sum rules \cite{Teryaev:2005uj,Anikin:2007yh}
and calculated related D-terms \cite{Polyakov:1999gs}. In this way GPDs became known
in the full region of $\xi$, and this allowed us to apply the inverse Radon transform
(derived for GPDs in \cite{Teryaev:2001qm}) to obtain double
distribution (DD) functions. Later on, we investigated subtleties of
applying the recently suggested prescription \cite{Radyushkin:2011dh} to
photon DDs.
We also turned to dual parametrization \cite{Moiseeva:2008qd}, where the partonic structure is
represented as an infinite series of t-channel exchanges, and
derived the quintessence functions.
\section{Photon GPDs in physical and nonphysical regions}
\label{AC}
DVCS amplitude can be represented as the tensorial decomposition \cite{Budnev:1974de}
\begin{equation}
T^{\mu\nu\alpha\beta} (\Delta_{T}=0) = \frac{1}{4}g^{\mu\nu}_Tg^{\alpha\beta}_T A_1+
\frac{1}{8}\left(g^{\mu\alpha}_Tg^{\nu\beta}_T
+g^{\nu\alpha}_Tg^{\mu\beta}_T -g^{\mu\nu}_Tg^{\alpha\beta}_T \right)A_2
+ \frac{1}{4}\left(g^{\mu\alpha}_Tg^{\nu\beta}_T - g^{\mu\beta}_Tg^{\alpha\nu}_T\right)A_3\,\label{tensorial}
\end{equation}

Amplitudes are presented as convolutions:
\begin{equation}
A_1(\xi)=\int\limits^{1}_{-1}dx C_V(x,\xi) H_1(x,\xi,0),~~~~~~A_3(\xi)=\int\limits^{1}_{-1}dx C_A(x,\xi) H_3(x,\xi,0)\label{W_H}
\end{equation}
with
\begin{equation}
C^q_{V/A}(x,\xi)=-2e_q^2\left(\frac{1}{x-\xi+i \eta}\pm\frac{1}{x+\xi-i \eta}\right)
\end{equation}

Dropping the factors $\frac{N_{C}e_q^2}{4\pi^2}\ln{\frac{Q^2}{m^2}}$ (as we are interested in $x,\xi$ dependencies only), photon GPDs corresponding to respective amplitudes in (\ref{tensorial}) are \cite{Friot:2006mm}:
\begin{eqnarray}
H^{q}_{1}(x,\xi,0)=\theta(x-\xi)\frac{x^2+(1-x)^2-\xi ^2}{1-\xi ^2}+\nonumber\\
\theta(\xi-x)\theta(x+\xi)\frac{x (1-|\xi |)}{|\xi |
(|\xi|+1)}-\theta(-x-\xi)\frac{x^2+(1+x)^2-\xi ^2}{1-\xi ^2},
\label{dvcs1}
\end{eqnarray}

\begin{eqnarray}
H^{q}_{3}(x,\xi,0)=\theta(x-\xi)\frac{x^2-(1-x)^2-\xi ^2}{1-\xi ^2}-\nonumber\\
\theta(\xi-x)\theta(x+\xi)\frac{1-|\xi
|}{|\xi|+1}+\theta(-x-\xi)\frac{x^2-(1+x)^2-\xi ^2}{1-\xi ^2}.
\label{dvcs3}
\end{eqnarray}

GDAs of the photon are \cite{ElBeiyad:2008ss}:

\begin{eqnarray}
\Phi^{q}_{1}(z',\zeta',0)=\theta(z'-\zeta')\frac{\overline{z'}(2z'-\zeta')}{\overline{\zeta'}}+
\theta(z'-\overline{\zeta'})\frac{\overline{z'}(2z'-\overline{\zeta'})}{\zeta'}+\nonumber\\
\theta(\zeta'-z')\frac{z'(2z'-1-\zeta')}{\zeta'}+\theta(\overline{\zeta'}-z')\frac{z'(2z'-1-\overline{\zeta'})}{\overline{\zeta'}}
\label{gda1}
\end{eqnarray}

\begin{eqnarray}
\Phi^{q}_{3}(z',\zeta',0)=\theta(z'-\zeta')\frac{\overline{z'}\zeta'}{\overline{\zeta'}}-\theta(z'-\overline{\zeta'})\frac{\overline{z'}\overline{\zeta'}}{\zeta'}
-\theta(\zeta'-z')\frac{z'\overline{\zeta'}}{\zeta'}+\theta(\overline{\zeta'}-z')\frac{z' \zeta'}{\overline{\zeta'}}
\label{gda3}
\end{eqnarray}

For convenience we used a more symmetric way \cite{Teryaev:2001qm,Diehl:1998dk} to define the coordinates of the photon distribution amplitudes by the difference of their momenta:
\begin{eqnarray}
z'=\frac{1-z}{2},\nonumber\\
\zeta'=\frac{1-\zeta}{2},\label{z_change}
\end{eqnarray}
where $z',\zeta'$ - are variables used in \cite{ElBeiyad:2008ss}.

Photon GPDs obtained from the DVCS process are defined in the region
with $|\xi|<1$ and $|x|<1$. They can be extended \cite{Teryaev:2001qm} to the unphysical region of $|\xi|>1$, expressing GPD via GDA $\Phi(z,\zeta)$ in their physical region $-1<z<1, -1<\zeta<1$.

In adopted normalization \cite{Diehl:2003ny, Friot:2006mm, ElBeiyad:2008ss} the relation between GPD and GDA takes the form:
\begin{equation}
 H(x,\xi)=\frac{1}{2} sgn(\xi) \Phi(\frac{x}{\xi},\frac{1}{\xi})\label{H_Phi2}
\end{equation}

Applying (\ref{H_Phi2}) to (\ref{gda1}) and (\ref{gda3}) we get GPD $H^{q}_{1}(x,\xi)$ for $|\xi|>1$,
\begin{equation}
\begin{cases}
 \frac{x-x \xi }{\xi ^2+\xi } & -1<x<1 \\
 \frac{(x-\xi ) \left(2 x \xi +\xi ^2-1\right)}{\xi  \left(\xi ^2-1\right)} & x>1 \\
 \frac{(x+\xi ) \left(-2 x \xi +\xi ^2-1\right)}{\xi \left(\xi ^2-1\right)} &x<-1
\end{cases}
\end{equation}
and analogously for $H^{q}_{3}(x,\xi)$,
\begin{equation}
\begin{cases}
 \frac{\xi -1}{\xi +1} & -1<x<1 \\
 \frac{2 (x-\xi )}{\xi ^2-1} & x>1 \\
 -\frac{2 (x+\xi )}{\xi ^2-1} & x<-1
\end{cases}
\end{equation}

\begin{figure}[h]
\begin{minipage}{0.5\linewidth}
\includegraphics[width=1\linewidth]{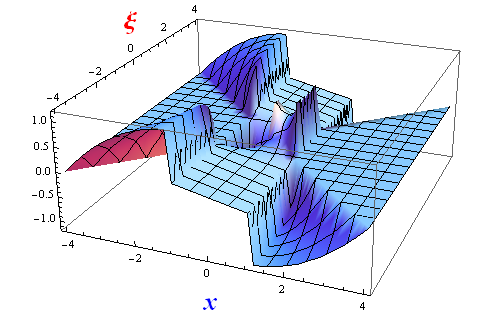}
\caption[]{\small $H^{q}_{1}(x,\xi)$}\label{gpd1_3d}
\end{minipage}
\hfill
\begin{minipage}{0.5\linewidth}
\includegraphics[width=1\linewidth]{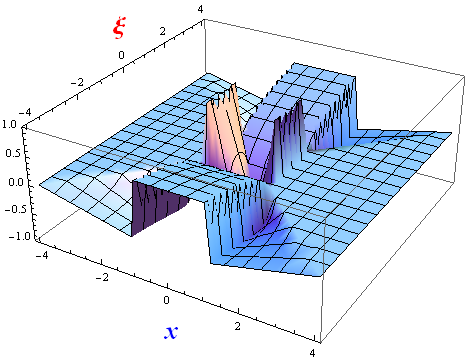}
\caption[]{\small $H^{q}_{3}(x,\xi)$}\label{gpd3_3d}
\end{minipage}
\end{figure}

As a result the functions are known in the full definition region ($|\xi|\leq 1,-1\leq x \leq 1$ and $|\xi|>1, -\xi \leq x \leq \xi$). Their 3D figures are illustrated in Figs. (\ref{gpd1_3d}) and (\ref{gpd3_3d}).
Their $\xi$ slices are illustrated in Figs.  (\ref{H1_2d}) and (\ref{H3_2d}) respectively.
As one can see the $x$-derivatives of $H_1(x,\xi,0)$ and $H_3(x,\xi,0)$ have discontinuities at the points $-1$, $-\xi$, $\xi$, $1$.
\textbf{
\begin{figure}[h]
\begin{minipage}{0.49\linewidth}\raggedleft
\includegraphics[width=1\linewidth]{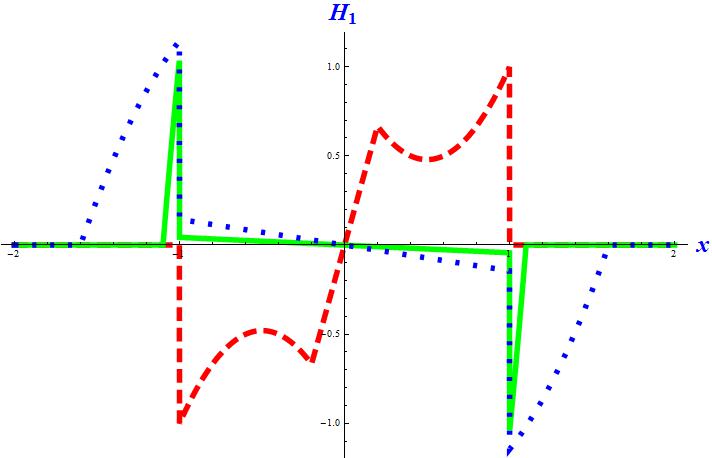}
\caption[l]{\small $H^{q}_{1}(x,\xi)$ for $\xi=0.2$ (dashed), 1.1(solid), 1.6 (dotted)}\label{H1_2d}
\end{minipage}
\hfill
\begin{minipage}{0.49\linewidth}\raggedleft
\includegraphics[width=1\linewidth]{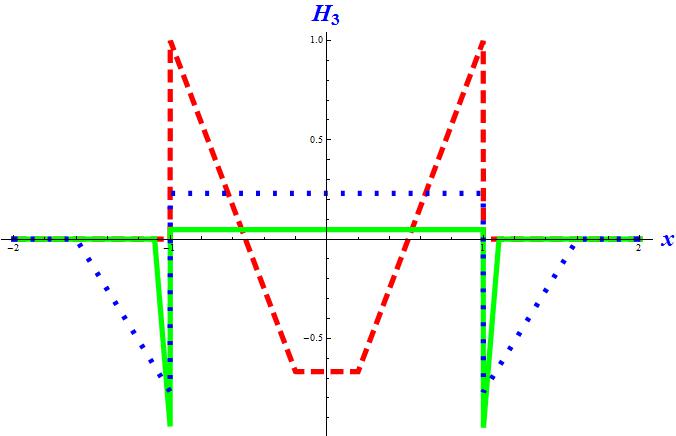}
\caption[]{\small $H^{q}_{3}(x,\xi)$ for $\xi=0.2$ (dashed), 1.1(solid), 1.6 (dotted)}\label{H3_2d}
\end{minipage}
\end{figure}}

\section{Holographic sum rules}

Here we investigate dispersion relations connecting real and imaginary parts of the DVCS amplitudes\cite{Teryaev:2005uj,Anikin:2007yh} in application to photon. The GPD contribution to hard exclusive amplitude (particularly for DVCS like in (\ref{W_H})) in the leading order is defined (dropping the $-2e_q^2$) through
\begin{eqnarray}
A_{1,3}(\xi,t)=\int\limits^{1}_{-1}dx H_{1,3}(x,\xi,t)[\frac{1}{x+\xi-i\epsilon}\pm\frac{1}{x-\xi+i\epsilon}]
\label{dvcs_amp}
\end{eqnarray}
or simply
\begin{equation}
A_{1,3}(\xi,t)=\int\limits^{1}_{-1}dx \frac{H_{1,3}(x,\xi,t)}{x+\xi-i\epsilon}
\end{equation}
\begin{equation}
Im A_{1,3}(x,t)=-\pi H_{1,3}(x,x,t).
\end{equation}

Note that the imaginary parts of the amplitudes $A_1$ and $A_3$ in the region $x>0$ differ only by the sign because
\begin{equation}H_1(x,x)=-H_3(x,x)=\frac{1-x}{1+x}\label{equal13}\end{equation}

The holographic sum rule for GPDs \cite{Teryaev:2005uj,Anikin:2007yh} guarantees that the relevant information of the function up to subtraction is contained in the 1-dimensional section:
\begin{eqnarray}
P\int\limits^{1}_{-1}\frac{H(x,\xi)-H(x,x)}{x-\xi}dx=\Delta=\int\limits^{1}_{-1}d\beta \int\limits^{1-|\beta|}_{-1+|\beta|}d\alpha\frac{G(\beta,\alpha)}{\alpha-1},
\label{hsr}
\end{eqnarray}
where $P\int$ means principal value of integral, while $\Delta$ does not depend on $\xi$, as a result,
\begin{eqnarray}
Re A(\xi)=\frac{P}{\pi}\int\limits^{1}_{-1}\frac{Im A(x)}{x-\xi}dx+\Delta
\label{hsr_imre}
\end{eqnarray}
Here and further we omit indices $1,3$ for amplitudes, GPDs and DDs when the equations are valid for both $H_1$ and $H_3$.
This equation tells us that the DVCS amplitude is defined through its imaginary part up to the subtraction constant. Let us stress once more, the information about DVCS in the leading order is contained on the line $x=\pm\xi$ manifesting its holographic property.

\begin{figure}[h!]
\centering
\begin{minipage}{0.7\linewidth}
\includegraphics[width=1\linewidth]{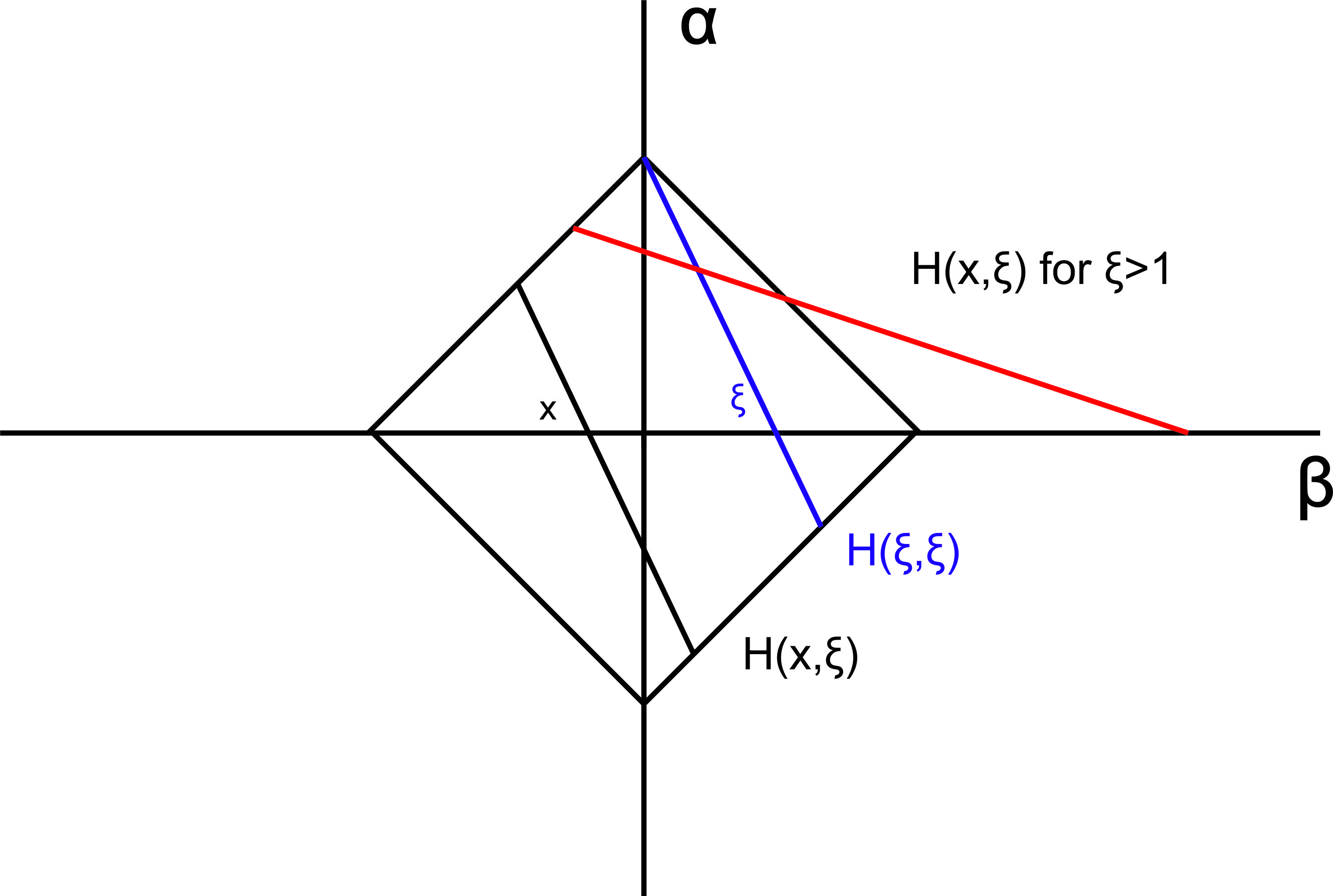}
\caption[]{\small $H(x,\xi)$ as integral of $F(\beta,\alpha)$}\label{dd_slices}
\end{minipage}
\end{figure}

\label{parsing_gpds}
To define $G$ one should consider GPDs connection to Double Distributions through integral \cite{Belitsky:2005qn}
\begin{eqnarray}
H(z,\xi)=\int\limits_{-1}^{1}d\alpha \int\limits_{-1+|\alpha|}^{1-|\alpha|}d\beta F(\beta,\alpha)\delta(z-\beta-\alpha\xi) \label{gpd_dd}
\end{eqnarray}
which can be interpreted \cite{Teryaev:2001qm} as the 2-dimensional Radon transform (RT) \cite{Radon:1917bf} (being the basis of tomography methods, for a recent review see e.g. \cite{Facchi:2010ct}), an integral of $F(\beta,\alpha)$ over the line with the slope $-1/\xi$ and crossing the $\beta$ axis at the point $x$. See Fig. \ref{dd_slices}.
The application inverse RT for GPD leads \cite{Teryaev:2001qm} to
\begin{equation}
F(\beta,\alpha)=-\frac{1}{2\pi^2}\int\limits_{-\infty}^{\infty}\frac{dz}{z^2}\int\limits_{-\infty}^{\infty}d\xi
(H(z+\beta+\alpha\xi,\xi)-H(\beta+\alpha\xi,\xi))
\end{equation}

There is an addition to the DD, the so called D-term \cite{Polyakov:1999gs, Teryaev:2001qm}, introduced in order to preserve the polynomiality condition which resides $|x|<|\xi|$,

\begin{equation}
H(z,\xi)=\int\limits_{-1}^{1}d\alpha \int\limits_{-1+|\alpha|}^{1-|\alpha|}d\beta (F(\beta,\alpha)+\xi \delta(\beta)D(\alpha))\delta(z-\beta-\alpha\xi).\label{gpd_via_d}
\end{equation}
More generally \cite{Teryaev:2001qm}
\begin{equation}
H(z,\xi)=\int\limits_{-1}^{1}d\alpha \int\limits_{-1+|\alpha|}^{1-|\alpha|}d\beta (F(\beta,\alpha)+\xi G(\beta,\alpha))\delta(z-\beta-\alpha\xi)\label{gpd_via_dd}
\end{equation}
for the two-DD representation, where $F$ and $G$ are defined ambiguously and can be reduced to the form \cite{Belitsky:2005qn}
\begin{eqnarray}
F(\beta,\alpha)&=&\beta f(\beta,\alpha)\\
G(\beta,\alpha)&=&\alpha f(\beta,\alpha),
\end{eqnarray}
where $f$ is defined in the single-DD representation:
\begin{equation}
\frac{H(z,\xi)}{z}=\int\limits_{-1}^{1}d\alpha \int\limits_{-1+|\alpha|}^{1-|\alpha|}d\beta f(\beta,\alpha)\delta(z-\beta-\alpha\xi)
\end{equation}

Let us find the D-terms of photon GPDs($H_1$ and $H_3$) exploring \cite{Teryaev:2010zz} that
\begin{equation}D(\alpha)=\Phi(\alpha,0),\end{equation}
as a result one gets
\begin{eqnarray}
D_{1}(\alpha)&=&(|\alpha|-1)(2|\alpha|+1)sgn(\alpha)\nonumber\\
\hfill
D_{3}(\alpha)&=&0
\label{D}
\end{eqnarray}

\begin{figure}[h!]
\begin{center}
\begin{minipage}{0.45\linewidth}
\includegraphics[width=1\linewidth]{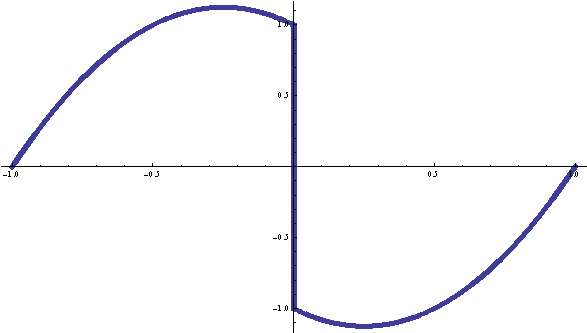}
\caption[]{ $D_{1}(\alpha)$ term}\label{figD1}
\end{minipage}
\end{center}
\end{figure}

The plot of $D_{1}(\alpha$) in Figs. (\ref{figD1}).
It is interesting the sign of the D-term is formally in accordance with stability criteria for nucleons in vacuum\cite{Polyakov:1999gs}, in nuclear matter \cite{Kim:2012ts} and Q-balls \cite{Mai:2012yc}.

Let us now check the holographic sum rules for photon GPDs (\ref{hsr}).
For $|\xi|<1$ integration interval divides into 3 intervals
\begin{equation}
\int\limits_{-1}^{1}\frac{H(x,\xi)-H(x,x)}{x-\xi}dx=[\int\limits_{-1}^{-\xi}+\int\limits_{-\xi}^{\xi}+
\int\limits_{\xi}^{1}]\frac{H(x,\xi)-H(x,x)}{x-\xi}dx,
\end{equation}
which for $H_1$ and $H_3$ respectively gives:
\tiny
\begin{eqnarray}
\left\{
\begin{aligned}
\int\limits_{-1}^{-\xi}\frac{H_{1}(x,\xi)-H_{1}(x,x)}{x-\xi}dx &=-\frac{\xi -2 \log (\xi +1)-1+\log (4)}{\xi -1}\\
\int\limits_{-\xi}^{\xi}\frac{H_{1}(x,\xi)-H_{1}(x,x)}{x-\xi}dx&=\frac{\xi  (\xi  (\log (4)-2)+4)-4 \log (\xi +1)-2+\log (4)}{\xi ^2-1}\\
\int\limits_{\xi}^{1}\frac{H_{1}(x,\xi)-H_{1}(x,x)}{x-\xi}dx&=\frac{3 \xi -2 \log (\xi +1)-1+\log (4)}{\xi +1}\\
\int\limits_{-1}^{-\xi}\frac{H_{3}(x,\xi)-H_{3}(x,x)}{x-\xi}dx &=\frac{2 \left(\xi  \log (\xi +1)-\xi  (1+\log (2))+\log \left(\frac{\xi +1}{2}\right)+1\right)}{\xi ^2-1}\\
\int\limits_{-\xi}^{\xi}\frac{H_{3}(x,\xi)-H_{3}(x,x)}{x-\xi}dx&=\frac{4 \xi  \log \left(\frac{2}{\xi +1}\right)}{\xi ^2-1}\\
\int\limits_{\xi}^{1}\frac{H_{3}(x,\xi)-H_{3}(x,x)}{x-\xi}dx&=\frac{2 \log (\xi +1)+2-\log (4)}{\xi +1}
\end{aligned}\label{ints}
\right.
\end{eqnarray}\normalsize

\begin{figure}[h]
\begin{minipage}{0.48\linewidth}
\includegraphics[width=1\linewidth]{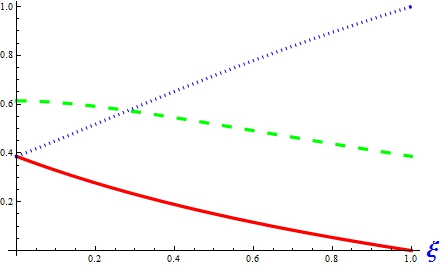}
\caption[]{\footnotesize{Contributions from integrals of $H_1$:$\int\limits_{-1}^{-\xi}$ - red solid;$\int\limits_{-\xi}^{\xi}$ - green dashed; $\int\limits_{\xi}^{1}$ - blue dotted}}\label{ints1}
\end{minipage}
\hfill
\begin{minipage}{0.48\linewidth}
\includegraphics[width=1\linewidth]{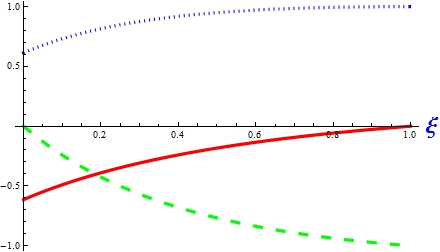}
\caption[]{\footnotesize{Contributions from integrals of $H_3$:$\int\limits_{-1}^{-\xi}$ - red solid;$\int\limits_{-\xi}^{\xi}$ - green dashed; $\int\limits_{\xi}^{1}$ - blue dotted}}\label{ints3}
\end{minipage}
\end{figure}

Contributions from these intervals exhibit dependence on $\xi$ as illustrated in Figs. \ref{ints1},\ref{ints3}, but their sum for $\xi \neq 0$ does not in accordance with holographic sum rule.

\begin{equation}\int\limits_{-1}^{1}\frac{H_{1}(x,\xi)-H_{1}(x,x)}{x-\xi}dx=2\ln 2 \text{ for $|\xi|<1$}\label{sub1low}\end{equation}
\begin{equation}\int\limits_{-1}^{1}\frac{H_{3}(x,\xi)-H_{3}(x,x)}{x-\xi}dx=0 \text{ for $|\xi|<1$}\label{sub3low}\end{equation}

The calculated values are in accordance with (\ref{hsr}) with r.h.s. (\ref{D}).
It is instructive to explore these integrals in detail. Expression in l.h.s. of
(\ref{hsr}) has a nontrivial limit $\xi \to 0$.
It happens that (cf. \cite{Radyushkin:2011dh}) this limit cannot be derived through substitution $\xi=0$ in (\ref{hsr}), because the central region $[-\xi,\xi]$ becomes strictly zero and so does the central integral while its limit (\ref{ints}) does not:
\begin{equation}\lim_{\xi \rightarrow 0}P\int\limits^{\xi}_{-\xi}\frac{H_1(x,\xi)-H_1(x,x)}{x-\xi}dx=2 - 2\ln 2\neq 0,\end{equation}
what leads to discontinuity at $\xi=0$.

In full analogy convolution integrals for $H_1$ and $H_3$ satisfy holographic sum rule in the region $|\xi|>1$ as well \cite{Teryaev:2010zz},

\begin{equation}\int\limits_{-\xi}^{\xi}\frac{H_{1}(x,\xi)-H_{1}(x,x)}{x-\xi}dx=2\ln 2 \text{ for $|\xi|>1$}\label{sub1hi}\end{equation}
\begin{equation}\int\limits_{-\xi}^{\xi}\frac{H_{3}(x,\xi)-H_{3}(x,x)}{x-\xi}dx=0 \text{ for $|\xi|>1$}\label{sub3hi}\end{equation}

As we can see, subtraction constants do not depend on $\xi$
\cite{Anikin:2007yh}, for any small but finite $\xi$.

\section{Recovering photon double distributions}

Different pairs $F(\beta,\alpha)$ and $G(\beta,\alpha)$ used in the two-DD representation (\ref{gpd_via_dd}) are connected through gauge transformations \cite{Teryaev:2001qm},
\begin{eqnarray}
F(\beta,\alpha)\rightarrow F(\beta,\alpha)+\frac{\partial \chi(\beta,\alpha)}{\partial \alpha}
\nonumber\\
G(\beta,\alpha)\rightarrow G(\beta,\alpha)-\frac{\partial \chi(\beta,\alpha)}{\partial \beta}
\label{gt}
\end{eqnarray}

Let us find the one transforming the D - term type DDs to
$F(\beta,\alpha)=\beta f(\beta,\alpha)$ and $G(\beta,\alpha)=\alpha f(\beta,\alpha)$ appropriate to the single-DD representation \cite{Belitsky:2005qn},

\begin{figure}[h!]\centerline{
\begin{minipage}[h]{0.3\linewidth}
$F(\beta,\alpha)=\beta f(\beta,\alpha)$
\vfill
$G(\beta,\alpha)=\alpha f(\beta,\alpha)$
\end{minipage}
\begin{minipage}[h]{0.2\linewidth}
$\chi(\beta,\alpha)$
\vfill
$\longleftrightarrow$
\end{minipage}
\begin{minipage}[h]{0.3\linewidth}
$F_{D}(\beta,\alpha)$
\vfill
$D(\alpha)$
\end{minipage}}
\end{figure}

$F_{D}$ corresponds to the $(F,G)$ pair, where $G(\beta,\alpha)=\delta(\beta) D(\alpha).$ $H_D(x,\xi)$ is the part of GPD corresponding to the D-term.

\begin{figure}[h!]
\begin{center}
\begin{minipage}{0.5\linewidth}
\includegraphics[width=1\linewidth]{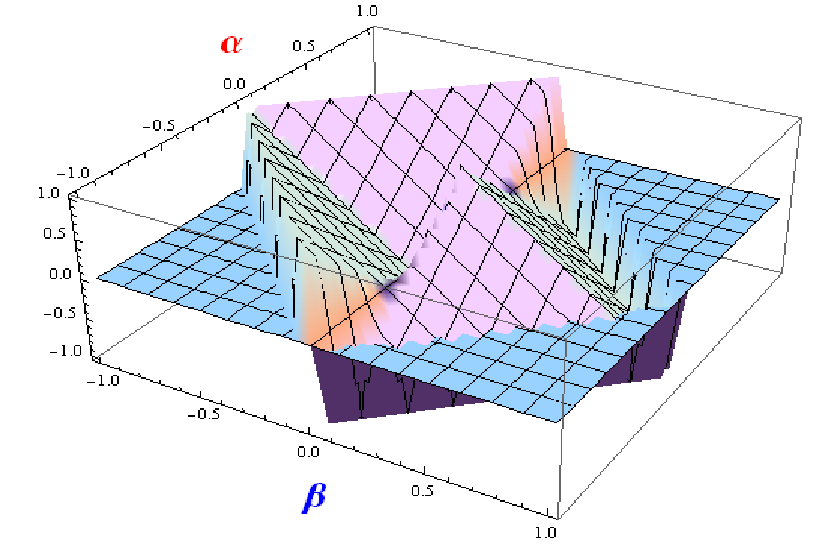}
\caption[]{Regular part of double distribution $F_{1D}(\beta,\alpha)$}\label{figF1D}
\end{minipage}
\end{center}
\end{figure}

Subtracting $H_D(x,\xi)=sgn(\xi)D(\frac{x}{\xi})$ from the $H(x,\xi)$ with the D-terms which we evaluated explicitly (\ref{D}), we proceeded with calculation of $F_{1D}(\beta,\alpha)$, $F_{3D}(\beta,\alpha)$ and $f_{1}(\beta,\alpha)$. Here we needed the inversion only for the single first term in the r.h.s. of (\ref{gpd_via_d}) which is much simpler. Firstly we made inverse Radon transform numerically, then conjectured analytic ansatz for DDs, and finally checked it analytically.
\begin{eqnarray}
F_{1D}(\beta,\alpha)=[2(1-|\beta|-|\alpha|)-1+\delta(\alpha)]sgn(\beta),
\label{F1D}
\end{eqnarray}
which is shown in Fig. (\ref{figF1D}).

Applying inverse Radon transform for $H_{3}(x,\xi)$ we obtain
\begin{eqnarray}
F_{3D}(\beta,\alpha)=\delta(\alpha)-1.
\label{F3}
\end{eqnarray}
Hence we derive $f_{1}(\beta,\alpha)$ using the inverse RT for $\frac{H_{1}(x,\xi)}{x}$ where one can identify the term $\delta(\beta)\frac{D_{1}(\alpha)}{\alpha}$
\begin{eqnarray}
f_{1}(\beta,\alpha)=\frac{\delta(\alpha)}{|\beta|}-1+2\delta(\beta)(1-|\alpha|)+\delta(\beta)\frac{D_{1}(\alpha)}{\alpha}
\label{f1}
\end{eqnarray}

\begin{figure}[h]
\begin{minipage}{0.5\linewidth}
\includegraphics[width=1\linewidth]{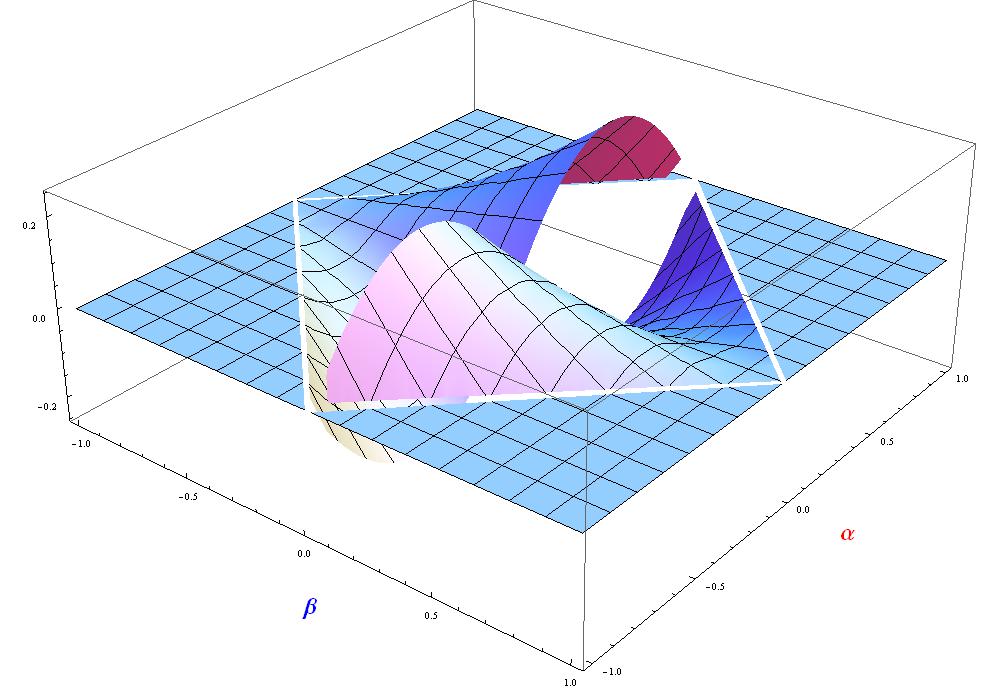}
\end{minipage}
\hfill
\begin{minipage}{0.6\linewidth}
\includegraphics[width=1\linewidth]{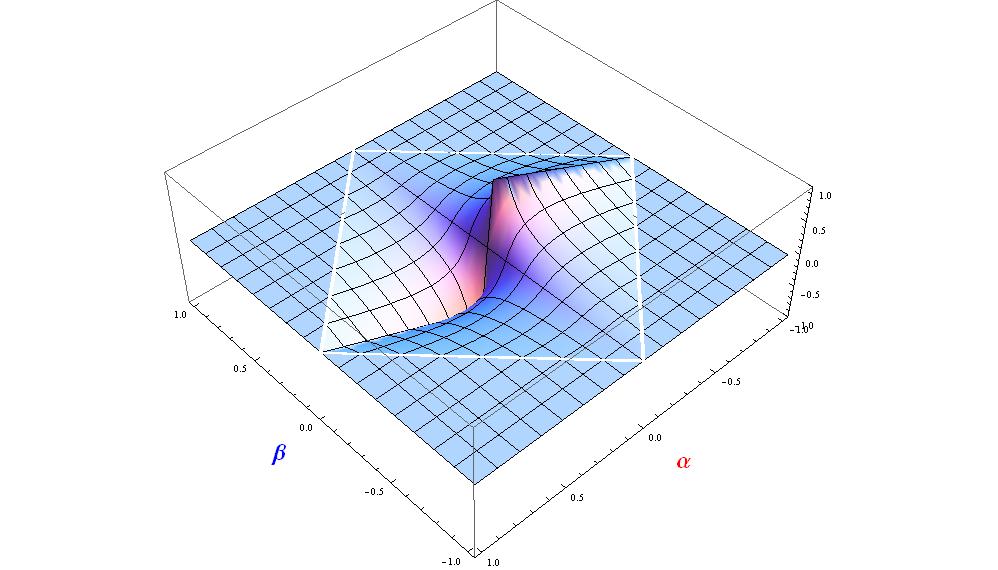}
\end{minipage}
\caption{Gauge transformation functions respectively $\chi_{1}(\beta,\alpha)$ and $\chi_{3}(\beta,\alpha)$}\label{figChi}
\end{figure}

This leads to
\begin{eqnarray}
F_{1}(\beta,\alpha)=\delta(\alpha) sgn(\beta)-\beta
\label{F1}
\end{eqnarray}
\begin{eqnarray}
G_{1}(\beta,\alpha)=-\alpha+2\delta(\beta)(1-|\alpha|)\alpha+\delta(\beta)D_{1}(\alpha)
\label{G1}
\end{eqnarray}
Now it is simple to derive the gauge transformation function (\ref{gt})
\begin{eqnarray}
\chi_{1}(\beta,\alpha)=\alpha sgn(\beta)(1 -|\beta|-|\alpha|)+C
\label{chi1}
\end{eqnarray}
Since $\chi_{1}$ is determined up to constant, we can set constant
$C$ to zero to make $\chi_{1}$ vanish at the border of the support
rhombus $|\beta|+|\alpha|=1$, after it $\chi_1$ is shown on the fig (\ref{figChi}).
Note that $\chi_{1}$ can also be calculated using only the $G_1$ function\cite{Teryaev:2001qm}
\begin{eqnarray}
\chi_{1}(\beta,\alpha)=\theta(\beta<0)\int\limits^{\beta}_{|\alpha|-1}d t G_1(t,\alpha)-\theta(\beta>0)\int\limits^{1-|\alpha|}_{\beta}d t G_1(t,\alpha)
\label{chi1_alt}
\end{eqnarray}
Although $D_3(\alpha)=0$, one may still perform the gauge transformation and write down the single-DD  representation generating fictitious (genuine but not necessary because it can be completely eliminated by gauge transform) DD $G_3$,
\begin{eqnarray}
\left\{
\begin{aligned}
\beta f_3(\beta,\alpha) -F_{3D}(\beta,\alpha) &= \frac{\partial \chi_3(\beta,\alpha)}{\partial\alpha}\\
0-\alpha f_3(\beta,\alpha)&= \frac{\partial \chi_3(\beta,\alpha)}{\partial\beta}
\end{aligned}
\right.
\end{eqnarray}
\begin{eqnarray}
-\alpha F_{3D}(\beta,\alpha)=\alpha \frac{\partial \chi_3(\beta,\alpha)}{\partial\beta}+\beta \frac{\partial \chi_3(\beta,\alpha)}{\partial\alpha}
\end{eqnarray}
\begin{eqnarray}
\chi_3(\beta,\alpha)=\int\limits^{1}_{-\infty}(-\alpha)F_{3D}(t\beta,t\alpha) d t
\end{eqnarray}
\begin{eqnarray}
\chi_3(\beta,\alpha)=\alpha(1-\frac{1}{|\beta|+|\alpha|}),
\end{eqnarray}
shown in Fig.  (\ref{figChi}). Consequently,
\begin{eqnarray}
f_3(\beta,\alpha)=\frac{\delta(\alpha)}{\beta}-\frac{sgn(\beta)}{(|\beta|+|\alpha|)^2}
\label{f3}
\end{eqnarray}

Now let us consider the plus-prescription notation \cite{Radyushkin:2011dh} for single-DD.
Plus-prescription is a standard method for regularization of singularities for parton densities. For GPD it can be used to select part not containing D-term. The aim of this construction is to make $\beta$-integrals of $f(\beta,\alpha)$ finite. Note that plus-prescripted function is a distribution which is defined by equation:
\begin{eqnarray}
\int\limits_{-1+|\alpha|}^{1-|\alpha|}\left[f(\beta,\alpha)\right]_{+} (K(\beta,\alpha))d \beta=
\int\limits_{-1+|\alpha|}^{1-|\alpha|}f(\beta,\alpha)(K(\beta,\alpha) - K(0,\alpha))d \beta
\label{plus_rule}
\end{eqnarray}
More formally, the general definition of DD "plus" term (presented by equation (75) in \cite{Radyushkin:2011dh}) is:
\begin{eqnarray}
\left[f(\beta,\alpha)\right]_{+} = f(\beta,\alpha)-\delta(\beta)\int\limits_{-1+|\alpha|}^{1-|\alpha|}f(\gamma,\alpha)d\gamma\label{rad_f75}.
\end{eqnarray}
Surely the integral in the r.h.s. diverges (its divergence is in fact the reason to use plus-prescription).
We will show that the application of this definition leading to equation
(74) in \cite{Radyushkin:2011dh}:
\begin{equation}
f(\beta,\alpha)=\left[f(\beta,\alpha)\right]_++\delta(\beta)\frac{D(\alpha)}{\alpha}
\end{equation}
is incorrect in case of photon single-DD due to presence $\delta(\alpha).$

For our known $f_1(\beta,\alpha)$ general definition leads to
\begin{equation}
\left[f(\beta,\alpha)\right]_{+} =\delta(\alpha)\left[\frac{1}{|\beta|}\right]_{+}-1+2\delta(\beta)(1-|\alpha|)
\end{equation}

At the same time from the single-DD approach it follows:
\begin{equation}
D_1(\alpha)=\alpha\int\limits^{1-|\alpha|}_{-1+|\alpha|}f_1(\beta,\alpha)d\beta
\label{d_from_dd}
\end{equation}
Verifying it for photon $f_{1}(\beta,\alpha)$ we see that it is correct. However, dividing it by $\alpha$ we will come to
\begin{equation}
\frac{D_1(\alpha)}{\alpha}\neq\int\limits^{1-|\alpha|}_{-1+|\alpha|}f_1(\beta,\alpha)d\beta
\label{wrong_d_alpha}
\end{equation}
because of the presence of $\delta(\alpha)$ in $f_1(\beta,\alpha)$ (\ref{f1}). Indeed, its integration gives
\begin{equation}
\int\limits^{1-|\alpha|}_{-1+|\alpha|}f_1(\beta,\alpha)d\beta=
\frac{D_1(\alpha)}{\alpha}+\delta(\alpha)\int\limits^{1-|\alpha|}_{-1+|\alpha|}\frac{d\gamma}{|\gamma|}
\end{equation}
where infinite discrepancy arises from the term containing $\delta(\alpha)$, which does not contribute in (\ref{d_from_dd}).
\section{Quintessence function}

\begin{figure}[h]
\begin{center}
\includegraphics[width=0.6\linewidth]{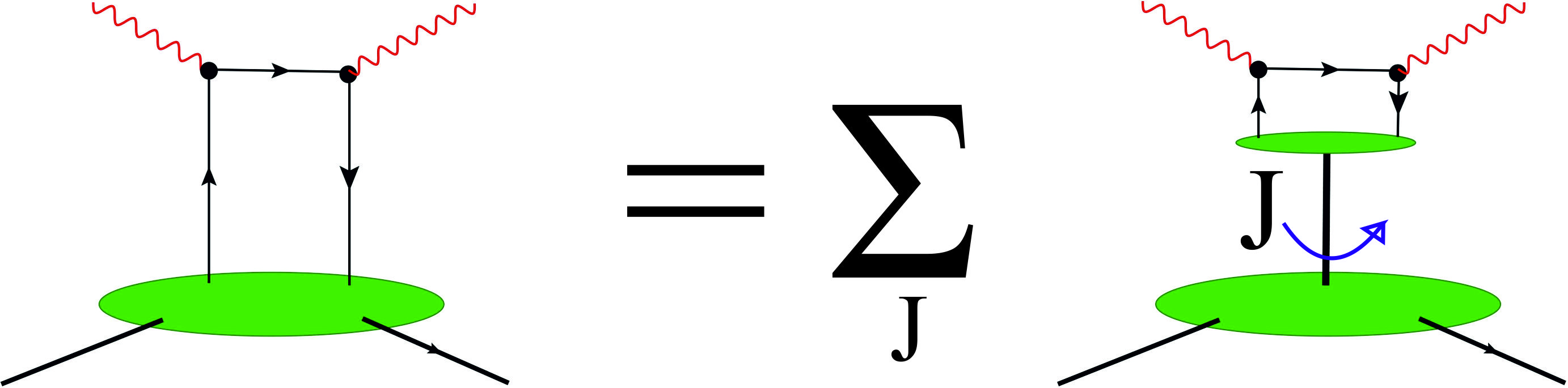}
\caption[]{GPD decomposition on t-channel exchanges with momentum $J$}\label{fig_N_md}
\end{center}
\end{figure}

\begin{figure}[h]
\begin{center}
\includegraphics[width=0.5\linewidth]{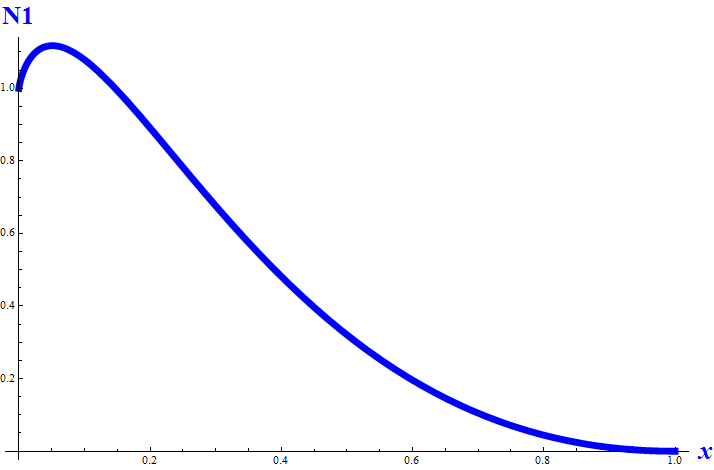}
\caption[]{Quintessence $N_1(x)$}\label{fig_N1}
\end{center}
\end{figure}

An alternative way to introduce GPDs instead of Double Distributions is the dual parametrization which represents parton distributions as an infinite series of t-channel exchanges\cite{Polyakov:2002wz}.
In the dual parametrization the amplitude is defined through the integrals \cite{Moiseeva:2008qd}:
\begin{eqnarray}
Im A(\xi,t) &=& \int\limits^{1}_{\frac{1-\sqrt{1-\xi^2}}{\xi}}\frac{dx}{x}N(x,t)
[\frac{1}{\sqrt{\frac{2x}{\xi}-x^2-1}}]\label{im_A_N}\\
Re A(\xi,t) &=& \int\limits^{\frac{1-\sqrt{1-\xi^2}}{\xi}}_{0}\frac{dx}{x}N(x,t)
[\frac{1}{\sqrt{1-\frac{2x}{\xi}+x^2}}+\frac{1}{\sqrt{1+\frac{2x}{\xi}+x^2}}-\frac{2}{\sqrt{1+x^2}}]\nonumber\\
+\int\limits^{1}_{\frac{1-\sqrt{1-\xi^2}}{\xi}}\frac{dx}{x} \! \! \!&N& \! \! \!(x,t)
[\frac{1}{\sqrt{1+\frac{2x}{\xi}+x^2}}-\frac{2}{\sqrt{1+x^2}}]+2D(t)\label{re_A_N}
\end{eqnarray}
which can be uniquely inverted in the following way \cite{Moiseeva:2008qd}:
\begin{eqnarray}
N(x,t)=\frac{2}{\pi}\frac{x(1-x^2)}{(1+x^2)^{3/2}}\int\limits^{1}_{\frac{2x}{1+x^2}}\frac{d\xi}{\xi^{3/2}} \frac{1}{\sqrt{\xi-\frac{2x}{1+x^2}}}\{\frac{1}{2}Im A(\xi,t)-\xi\frac{d}{d\xi}Im A(\xi,t)\}\label{N_def}
\end{eqnarray}
Its physics content is more clear for Mellin moments of $N(x,t)$ which are directly related to the contributions of quark-antiquark states with definite angular momentum in the t-channel\cite{Polyakov:2007rv}:
\begin{equation}
\int\limits^{1}_{0} d x x^{J-1} N(x,t)=\frac{1}{2} \int\limits^{1}_{-1} d z \frac{\Phi_{J}(z,t)}{1-z},
\end{equation}
where $\Phi_{J}(z,t)$ is the distribution amplitude of exchange with the angular momentum $J$ as illustrated in Fig. \ref{fig_N_md}.
Substituting $\pi H(\xi,\xi)$ we obtain quintessence for the photon
\begin{equation}
N_1(x)=-N_3(x)=\frac{(x-1)(x^2+2x \log(x)-1)}{(x+1)^2},
\label{N1}
\end{equation}
illustrated in Fig. (\ref{fig_N1}).
As we see, $N_1(x)=-N_3(x)$ because likewise do the imaginary parts of the respective amplitudes, see (see \ref{equal13})

\begin{figure}[h!]
\begin{center}
\includegraphics[width=0.5\linewidth]{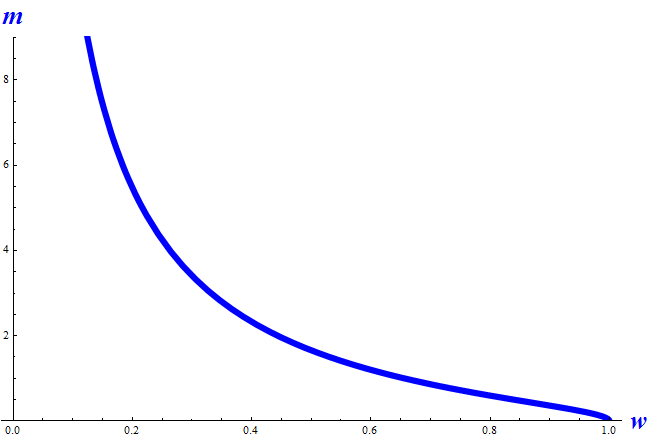}
\caption[]{ $m(\omega)$}\label{fig_m_w}
\end{center}
\end{figure}

\begin{figure}[h!]
\begin{center}
\includegraphics[width=0.5\linewidth]{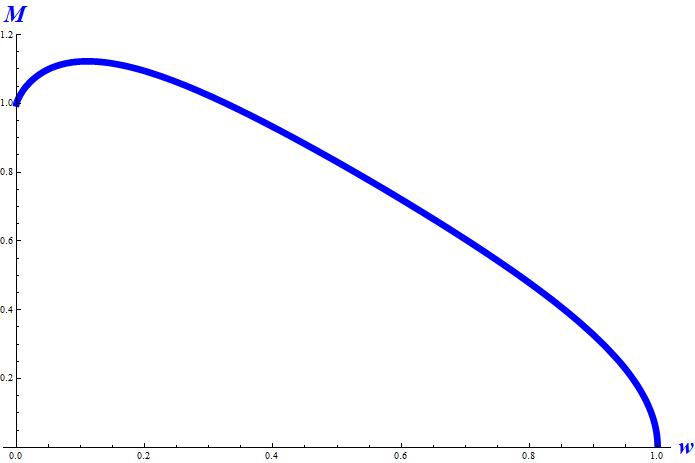}
\caption[]{ $M(\omega)$}\label{mw_w}
\end{center}
\end{figure}

Equation (\ref{im_A_N}) (after substitution $\frac{1}{w}=\frac{1}{2}(1+\frac{1}{x})$ which is connected with Joukowskii conformal map) is exactly the Abel transform \cite{Moiseeva:2008qd}. This means that it can be interpreted as a slice of 2-dimensional axially symmetric function \begin{equation}m(\omega)=\frac{M(\omega)}{\omega}\end{equation}

\begin{equation}
Im A(\xi,t)=\int\limits^{1}_{\xi}\frac{d\omega}{\omega}M(\omega,t)\frac{\sqrt{\xi}}{\sqrt{\omega-\xi}}
\end{equation}
For the photon
\begin{equation}
m(\omega)=\frac{\sqrt{1-\omega^2}+\omega \tanh^{-1}(\sqrt{1-\omega^2})}{\omega(\omega+1)^{3/2}},
\end{equation}
illustrated in Fig. \ref{fig_m_w}, and $M(\omega)=\omega m(\omega)$ in Fig. \ref{mw_w}

\section{Conclusion}

Dispersion relations for hard exclusive amplitudes are highly useful tools for QCD. In this sense, it seems extremely important to verify them. However, it is a rather difficult to test them for hadrons. That is why the photon is the unique object to analytically check the dispersion relations and related QCD tools. Also, the expressions of photon GPDs can give a hint for constructing hadron GPDs. At the same time the possibility of a direct experimental examination of photon GPDs should not be ruled out. One of the processes which seems to be a possible application is the photon splitting in the nucleus electromagnetic field (see e.g. \cite{Kaufhold:2005vj}) when the outgoing photons are collinear to the initial.

In this work we extended the original photon GPDs to the unphysical region of $|\xi|>1$ using their GDAs. Having completely defined GPDs we verified the holographic sum rule (\ref{hsr}) for the DVCS amplitudes in the leading order (\ref{hsr_imre}). We calculated the D-terms (\ref{D}) and subtraction constants (\ref{sub1low}),(\ref{sub3low}). Also, we derived photon DDs for the single-DD (\ref{f1}),(\ref{f3}) and two-DD (\ref{F1D}),(\ref{F3}) approaches with the help of the inverse Radon transform. Afterwards, we verified equations for the plus-prescription and noted that the various definitions of the D-term via the single-DD function are not equivalent. Later, using the dual parametrization approach we calculated quintessence functions (\ref{N1}) for photon GPDs. As an outlook let us mention the possibility to generalize the suggested approach to impact parameter dependent Photon GPDs \cite{Mukherjee:2011an, Mukherjee:2011bn}.

\section{Acknowledgments}
\label{acknowledgments}
The authors would like to thank I. Anikin, B. Pasquini, B. Pire, M. Polyakov, L. Szymanowski, P. Schweitzer and especially A. Radyushkin for helpful discussions and comments.
The work is supported in part by RFBR grants 12-02-00613, 11-02-01538, 11-02-01454, Heisenberg-Landau and Votruba-Blokhintsev programs.


\newpage
\begin{thebibliography}{99}

\bibitem{Belitsky:2005qn}
A. Belitsky and A. Radyushkin, Phys.Rept. {\bf 418},  1  (2005), dedicated to
  Anatoly V. Efremov on occasion of his 70th anniversary.

\bibitem{Diehl:2003ny}
M. Diehl, Phys.Rept. {\bf 388},  41  (2003), habilitation thesis.

\bibitem{Goeke:2001tz}
K. Goeke, M.~V. Polyakov, and M. Vanderhaeghen, Prog.Part.Nucl.Phys. {\bf 47},
  401  (2001).

\bibitem{Friot:2006mm}
S. Friot, B. Pire, and L. Szymanowski, Phys.Lett. {\bf B645},  153  (2007).

\bibitem{ElBeiyad:2008ss}
M. El~Beiyad, B. Pire, L. Szymanowski, and S. Wallon, Phys.Rev. {\bf D78},
  034009  (2008).

\bibitem{Witten:1977ju}
E. Witten, Nucl.Phys. {\bf B120},  189  (1977).

\bibitem{Mukherjee:2011bn}
A. Mukherjee and S. Nair, Phys.Lett. {\bf B706},  77  (2011).

\bibitem{Mukherjee:2011an}
A. Mukherjee and S. Nair, Phys.Lett. {\bf B707},  99  (2012).

\bibitem{Teryaev:2001qm}
O. Teryaev, Phys.Lett. {\bf B510},  125  (2001).

\bibitem{Teryaev:2005uj}
O. Teryaev, arXiv:hep-ph/0510031  (2005).

\bibitem{Anikin:2007yh}
I. Anikin and O. Teryaev, Phys.Rev. {\bf D76},  056007  (2007).

\bibitem{Polyakov:1999gs}
M.~V. Polyakov and C. Weiss, Phys.Rev. {\bf D60},  114017  (1999).

\bibitem{Radyushkin:2011dh}
A. Radyushkin, Phys.Rev. {\bf D83},  076006  (2011).

\bibitem{Moiseeva:2008qd}
A.~M. Moiseeva and M.~V. Polyakov, Nucl.Phys. {\bf B832},  241  (2010).

\bibitem{Budnev:1974de}
V. Budnev, I. Ginzburg, G. Meledin, and V. Serbo, Phys.Rept. {\bf 15},  181
  (1975).

\bibitem{Diehl:1998dk}
M. Diehl, T. Gousset, B. Pire, and O. Teryaev, Phys.Rev.Lett. {\bf 81},  1782
  (1998).

\bibitem{Radon:1917bf}
J. Radon, Ber. Verh. Sachs. Akad. Wiss. Leipzig, Math.-Nat. Kl. 69 (1917)
  262-277  .

\bibitem{Facchi:2010ct}
P. Facchi and M. Ligabo, AIP Conference Proceedings {\bf 1260},  3  (2010).

\bibitem{Teryaev:2010zz}
O. Teryaev, PoS {\bf DIS2010},  250  (2010).

\bibitem{Kim:2012ts}
H.-C. Kim, P. Schweitzer, and U. Yakhshiev, arXiv:1205.5228  (2012).

\bibitem{Mai:2012yc}
M. Mai and P. Schweitzer, arXiv:1206.2632  (2012).

\bibitem{Polyakov:2002wz}
M. Polyakov and A. Shuvaev, hep-ph/0207153  (2002).

\bibitem{Polyakov:2007rv}
M. Polyakov, Phys.Lett. {\bf B659},  542  (2008).

\bibitem{Kaufhold:2005vj}
C. Kaufhold and F.~R. Klinkhamer, Nucl.Phys. {\bf B734},  1  (2006).

\end{thebibliography}
\end{document}